\documentclass{ws-procs961x669}            
\begin{document}
\title{The Force and Gravity of Events}

\author{Robert Delbourgo$^*$}

\address{School of Physical Sciences, University of Tasmania,\\
Hobart, Tasmania 7001, Australia\\
$^*$E-mail: bob.delbourgo@utas.edu.au}

\begin{abstract}
Local events are characterized by ÒwhereÓ, ÒwhenÓ and ÒwhatÓ. Just as (bosonic) spacetime forms the backdrop for location and time, (fermionic) property space can serve as the backdrop for the attributes of a system. With such a scenario I shall describe a scheme that is capable of unifying gravitation and the other forces of nature. The generalized metric contains the curvature of spacetime and property separately, with the gauge fields linking the bosonic and fermionic arenas. The super-Ricci scalar can then automatically yield the spacetime Lagrangian of gravitation and the standard model (plus a cosmological constant) upon integration over property coordinates.
\end{abstract}


\bodymatter

\section{Life with Salam: Personal Reminiscences }
This conference commemorates the life and achievements of Abdus Salam. It is
therefore incumbent upon me to begin by drawing a picture of the man with a few private 
reminiscences that may convey something about his greatness, genius and 
humanity. I know that {\em some} of you have worked with him in {\em some} capacity at 
{\em some} stage, but I suspect that only very few of you will have had the privilege of
interacting 17 years with Abdus Salam as I have done: first as an undergraduate, then 
as a postgraduate, postdoc, and eventually as an academic colleague and scientific 
collaborator. I shall cover the period 1959-1976 when I was closely involved with 
him; there are a few others present here who can competently fill in the later 
years to leave you with a more complete portrait of Salam. If certain members of this 
audience have heard my vignettes of him before, I apologize in advance, but with these 
reminiscences you may at least enjoy reliving fond memories of him.

Salam was a man in a hurry; his reputation preceded him everywhere. As a lowly
undergraduate student I first came across him in 1959 when we had to choose our
third year specialty by making a selective tour of the various research departments
at Imperial College. At that time Salam was housed in the Mathematics section
(before moving to Physics in 1960) and I can vividly remember his verve and 
vivacity as he explained to us his latest pet project, which happened
to be chiral symmetry and gamma-5 invariance for favouring massless left-handed
neutrinos and leading to parity violation. That discourse went right over all our heads 
at the time but they coaxed me at least to turn to theoretical physics for my final year
specialty. I am sure others have succumbed to Salam's persuasive abilities
on whatever topic he expounded. That very year he taught us advanced quantum
mechanics a-la-Dirac, whom he idolized; his lectures seemed pretty good to me,
so in 1960 I embarked on my PhD, with Salam acting as my supervisor from 1961
onwards. During those years Salam's areas of interest were on vanishing of renormalization
constants for composite systems, Lie Groups and on the ``Gauge Technique''. He took a 
keen interest in my research topic and would enquire every morning as to what progress 
I had made -- putting great pressure on me, as he did on all his other students. He was
always bubbling with new ideas and postgraduates found it very hard to survive his changes 
of tack or emphasis; but it definitely steeled us. 

That period saw the development of the eightfold way and I very well remember Salam's heated arguments with Neeman that emanated from his office across the corridor. Salam
lost out on the birth of SU(3) because of his insistence on a fundamental (Sakata) triplet
so you can say that he tripped up on that. However in hindsight was he that far out? 
Think quarks and you will agree that his intuition was amazing. Yes, he could make mistakes
--- and who does not --- but on most things his inspiration was spot on. When I sometimes
asked him where and how he got his latest idea, he would give a wicked smile and point
upwards. He always moved on to something new when an old idea was established
and played out; he was never one for pot-boilers and he never suffered fools gladly, 
in private anyway.  When he became somewhat contemptuous of the work of some scientists
he referred to them as `tom-tits' or `broken reeds' or `youths'.  However in public he was
always polite and he encouraged anyone who presented a new concept. If there is
one lesson that Salam has taught me it is that one should not be ashamed to
move on if a concept is not bearing fruit. That may explain why there was always
great anticipation whenever he delivered a lecture on some topic: the expectation was
that he would spring something new on the audience.

Salam was a demon for hard work. For instance, in the summer of 1967 he had an
appendectomy; I visited him in hospital two days after his operation and it was not long
before he launched into discussing multiquark states and their current algebras, despite
his obvious physical discomfort. He was well travelled and especially during 1962-65
when in the process of setting up ICTP; funding problems beset him for a good while
and he would rile at politicians who opposed his initiative, including the Australian 
representatives in UNESCO!  Owing to his regrettable experiences in Pakistan 
after leaving Cambridge and his constant bemoaning of the decline of Islamic
science, he felt a driving need to found such a centre to assist third world countries. He
had a special affinity for isolated scientific personnel who, like himself at first, struggled
to keep abreast of the latest advances. He believed that initially concentrating on 
theoretical studies would serve the purpose as it would cost relatively little but represented 
the forefront of physics; later on the Centre could be used as a launchpad for other
branches of science. At Imperial College, and later at Trieste, Salam became a major
magnet for Pakistani students as well as those from Africa and Latin America.
The place was abuzz with them and Salam took great pains to foster their work.
His initiatives and his constant movement gave him little time for relaxation and I vividly 
remember several meetings that John Strathdee and I had with him in the hotel lobby at 
Trieste, after his energy-sapping perambulations. I think he was able to maintain his stamina
because he was an early riser and went to bed early too. My request that he not ring 
me before 7am, unless there was an emergency, must have tested his self-control.

Always the perfect host, he warmly welcomed new visitors to IC and ICTP by inviting
them to dinner at his home or elsewhere. He was not in the least pretentious about 
the venue. I recall one occasion at IC when Bruno Zumino came to give us a lecture. 
Instead of taking him to the Staff Club for lunch he opted for the College cafeteria so he 
could mingle with the `plebs' and sample the canteen fare, which he rather savoured!
The thing that most impressed John and I about his eating habits was when we 
were consuming fish; Salam would crunch his way through the spine and bones, leaving
only the head and tail! If waiters were tardy with producing the bill, Salam would get up 
and leave the `trattoria' when his patience ran out; the sight of the `camerieri' scurrying 
after him with `il conto' was pure comedy.  
At coffee he would often ruminate about the heyday of Arabic science and how vital it was
 in the Middle Ages for passing on the Greek scientific legacy to Europe via Spain.
 
 I have been asked by the organizers to comment upon the the birth of
 the standard model during 1967 and Salam's prominent role in it.
 This is an excellent occasion to set the record straight and recount my view of its
 history; if nothing else to refute innuendos which have occasionally surfaced during 
 the 1970s that Salam was not deserving of the Nobel Prize.
 That autumn of 1967 I had been in charge of organizing the seminars at
 IC. Because Salam was constantly on the move and hardly spent more than one month
 at a stretch in London, I arranged with him to give a couple of lectures on his
recent research (in October, to the best of my recollection) during his spell at IC
to kick off the seminar season, as it was early in the academic year. He agreed
to do so even though the audience attending those talks was somewhat thin. Paul Matthews 
was certainly present, but Tom Kibble was away in sabbatical in the USA. My memory 
of his lectures is a bit indistinct nowadays, but I do remember that he kept on
invoking these k-meson tadpoles which disappeared into the vacuum which induced
the spontaneous breaking of the gauge symmetry: what we now know as the expectation 
value of the Higgs boson. The resulting model looked rather ugly -- and it still is -- and I 
admit that I paid little attention to it; nor do I think that Salam himself was especially 
enraptured by the model's beauty. A week or so later, I wandered into the Physics Library
and came across Steven Weinberg's Physical Review Letter, which I noticed looked
suspiciously like Salam's attempt. I showed the article to Salam, who was rather
troubled that it was almost the same as his own research, but which was of course entirely
{\em independent}.  Matthews and I urged him to publish his work at the earliest opportunity
and this happened to be the upcoming Nobel Symposium. As they say, ``the rest is history''.
I hope that this account of the events at the time scotches all aspersions that Salam should not
have been a prize recipient.

Above all Salam impressed upon us the importance of tackling challenging problems: to
prospect for new scientific fields and abandon raking over old coals, if one was to make one's
mark.. He lived up to that precept throughout his life, in spite of accusations that he had a
scattergun approach to physics. It is a lesson that some young scientists today should heed. 
In his last years, despite his grave illness, he addressed the puzzle as to why certain life forms have particular handedness; what could be more fundamental or significant than that? I know that I miss his wise words, his friendship, his guidance, his generosity and his humanity. This conference is therefore a personal acknowledgement of how much he helped to 
shape my own career. More widely, it is a timely reminder of how much the scientific 
landscape and international  developments owe to him. The ICTP is a permanent 
testament to that.

\section{An algebraic framework for events}
This brings me to the scientific part of my talk. The material which I will present
is sufficiently different from other attempts at unification of forces that I rather fancy 
Salam might have given it a nod of approval. Two years ago, at the Dyson 90th anniversary conference, I outlined\cite{RDDyson} how it is possible to unify gravity with the simplest of all
forces, electromagnetism -- Einstein's eldorado --  simply by appending a single complex
anti-commuting Lorentz {\em scalar} variable to spacetime, {\em not a spinor}; importantly
no infinite KK modes arise. My partner in crime (Paul Stack) and I have made considerable 
progress since then and I will now try to summarise how  to unify gravity with the other forces 
of nature through a relatively simple supermetric. Our attempts in this direction have been
motivated by the present parlous state of particle physics and the snail's pace of progress in
this area over the last 40 years. Here is a statement which may bring me some opprobium: namely, apart from the timely discovery of the Higgs boson, emergence of  multiquark states 
and significant astrophysical advances, there is very little to celebrate in our attempts to unravel nature at the most basic level. This is in spite of determined, quasi-herculean efforts of theorists who have persistently espoused/promoted very clever ideas. So far, Nature stubbornly refuses to cooperate by providing us with unequivocal experimental 
signs of SUSY, strings/branes and other ingenious proposals. 
It seems that the simple standard model of particles and cosmology still rules.
Nonetheless its plethora of parameters have spurred theorists to search for generalizations
of the standard model which may help to cut down the number of arbitrary constants and 
leave room for mysterious dark matter. Many schemes have been put forward. These
usually add other gauge fields, sterile particles, invoke enlarged groups and 
introduce scalar fields, perhaps associated with cosmological inflation. 
My feeling is that these ideas are very much hit-or-miss and they do seem to lack a 
fundamental basis. I think Salam might have looked askance at them. 
Anyhow here goes...

For many years we have become accustomed to the notion of spacetime events, with 
local fields (belonging to representations of some gauge group) interacting at a particular 
site and time. The $x=(t,{\bf x})$ spacetime continuum serves as the backdrop for 
the `when' and `where' of an event. But, until one specifies the fields involved in the
interaction, the `what' of the action is left open, to be determined by experiment. Now 
we should realise that any event necessarily consists of a transaction or a change of property 
at a location. (The transaction is usually communicated by a gauge field.)
It occurred to  me that it might be possible to provide a mathematical backdrop for 
`properties' or `attributes' of the participating fields by invoking  a property space 
with its own set of coordinates.  As far as we can tell there seem to be a finite number of quantum numbers or properties in nature. So the basic idea is to put some
mathematics into the `what' of the event by invoking anticommuting (Lorentz scalar) 
coordinates $\zeta$; these should serve to provide the setting for the gauge groups and 
particle attributes and fields should be functions of these $\zeta$ as well as
spacetime location $x$. The {\em full} action is to be integrated over the properties $\zeta$
like one does for $x$. The reason why I have picked $\zeta$ as anticommuting is because 
when an object is endowed by several such properties, the melange is necessarily finite; 
and since the square of a property vanishes it means that once a fundamental constituent
possesses that attribute it cannot doubly have it. Of course since we are dealing with 
quantum mechanics in the long run, these properties must be complex so the 
anti-attribute $\bar{\zeta}$ should be permitted. By combining properties with 
anti-properties one can build up `generations' of particles possessing the same 
{\em overall} attributes. In some sense $N$-extended supersymmetry is based
on the same idea but it suffers badly from spin state proliferation.

The question is how many property coordinates $\zeta$ are needed? There must 
be enough to  describe the visible world. The pioneers of unified forces\cite{GG,FM} have
forged the way and provided the inspiration. Despite some criticisms to which these full
gauge groups have been subjected, I have opted for SU(5) and SO(10) gauge models; 
these have many attractive features, so for now I will suppose that there are five 
independent\footnote{I have found that four $\zeta$ are definitely insufficient to produce 
three generations at least.} complex $\zeta$. Later on we will be forced to subtly enlarge 
this number in order to reflect the incontrovertible fact that fermions of distinct chiralities 
-- through their electroweak characteristics -- behave quite differently at low energy; 
thus experiment obliges us to distinguish between left and right properties.

\section{Mathematical Description}
By enlarging spacetime $x$ with $\zeta$ we hope to encompass {\em all} possible fundamental events. Even though the term has been overused we will assume that there exist `superfields' 
$\Phi(X)$ and $\Psi(X)$ which are are functions of the super-coordinate 
$X^M \equiv (x^m, \zeta^\mu,\zeta^{\bar{\mu}})$. 
The idea is that an integral over products of just one or two superfields can provide the 
entire action for every event. The calculus for handling the combination of 
bosonic $x$ and fermionic $\zeta$ is well-established\footnote{We developed this from 
scratch as we wanted to adhere to Einstein up notation for coordinates and traditional 
left operations like differentiation. Also we wanted to settle the notation to our own satisfaction. 
Hereafter, Latin letters signify spacetime and Greek letters signify property.
Early letters of the alphabet connote flatness while later letters imply curved space.} 
and the graded character of $X$ means that Berezinian integration is to be adopted for 
property integration, with super-determinants coming into play. By curving the superspace
we will automatically be able to describe gravity and the other forces of nature, as we
shall see. 

But let us start with flat space and assume parity conservation; presently we shall 
improve on this by adding gauge fields and parity violation. With five $\zeta$ we are dealing
with an overarching Sp(10) group. The supermetric distance for flat OSp(1,3/10) is
\begin{equation}
ds^2 = dx^a dx^b \eta_{ba}+\ell^2(d\zeta^{\bar{\alpha}} d\zeta^{\beta}\eta_{\beta\bar{\alpha}}
       + d\zeta^\alpha d\zeta^{\bar{\beta}} \eta_{\bar{\beta}\alpha}) /2 ,
\end{equation}
where $\eta_{ba}$ is Minkowskian and $\eta_{\beta\bar{\alpha}}=-\eta_{\bar{\alpha}\beta}=
{\delta_\beta}^\alpha$; also a fundamental length scale $\ell$ must be introduced because 
we are presuming that property $\zeta$ is dimensionless.

The Bose fields are to be associated with even powers of $\zeta$ and its conjugate, while
the Fermi fields are connected with odd powers. Let us reserve the labels 1,2,3 for colour 
property or `chromicity' and 0,4 to neutrinicity, electricity. The quantum numbers which are
ascribed to these, viz.
\begin{eqnarray}
{\rm Charge~}Q(\zeta^0,\zeta^{\bar{1}},\zeta^{\bar{2}},\zeta^{\bar{3}},\zeta^4) 
  & = &(0,1/3,1/3,1/3,-1)\\
{\rm Fermion~Number~}F(\zeta^0,\zeta^{\bar{1}},\zeta^{\bar{2}},\zeta^{\bar{3}},\zeta^4) 
  & = &(1,-1/3,-1/3,-1/3,1)
\end{eqnarray}
really only come to life when one introduces the gauge fields, as we soon will. Given the
assignments (2), the lepton doublet generations are connected with $(\zeta^0,\zeta^4)$, 
multiplied by powers of $\zeta^{\bar{\rho}}\zeta^\rho$; the quark generations arise
more subtly. 

Component fields $\phi$ and $\psi$ emerge\cite{DJW} when we expand $\Phi$ and $\Psi$ 
as polynomials in $\zeta~\&~\bar\zeta$.
Fermions are to be associated with odd powers and bosons with even powers of attributes.
Charge conjugation of course corresponds to the `reflection'  operation
$\zeta\leftrightarrow\bar{\zeta}$ and we may define a duality operation (that does not affect 
the SU(5) representations) under which 
$(\zeta)^r(\bar\zeta)^s\leftrightarrow(\zeta)^{5-s}(\bar\zeta)^{5-r}$. By imposing selfduality
{\em or} anti-selfduality on the superfields we can greatly reduce the number of independent 
component fields arising in the $\zeta$- expansion This is detailed in ref. 4.
In amongst the boson $\Phi$ states are nine colour neutral uncharged mesons of which
the combination 
$\zeta^4\zeta^{\bar{1}}\zeta^{\bar{2}}\zeta^{\bar{3}}$ is recognizable
as the standard model Higgs. However, the quark isomultiplets $\psi$ which exist in $\Psi$ 
are slightly different from the standard model!  The up- \& down- quarks come as two weak isodoublets/singlets and part of a weak {\em isotriplet/isodoublet/isosinglet} contained in 
SU(5) representations of dimension 45. Thus,
\[
\left( \begin{array}{c} 
        U^{[\bar{\mu}\bar{\nu}]}\sim \zeta^{\bar{\mu}} \zeta^{\bar{\nu}} \zeta^0\\ 
        D^{[\bar{\mu}\bar{\nu}]}\sim \zeta^{\bar{\mu}} \zeta^{\bar{\nu}} \zeta^4
        \end{array}   \right),
        \,\,
\left( \begin{array}{c}
        U'^{[\bar{\mu}\bar{\nu}]}\sim \zeta^{\bar{\mu}}\zeta^{\bar{\nu}}\zeta^0\zeta^{\bar{4}}\zeta^4\\ 
        D'^{[\bar{\mu}\bar{\nu}]}\sim \zeta^{\bar{\mu}}\zeta^{\bar{\nu}}\zeta^4\zeta^{\bar{0}}\zeta^0
        \end{array}   \right) ,
        \,\,
\left( \begin{array}{c}
        U''^{\lambda} \sim \zeta^{\lambda}\zeta^{\bar{4}}\zeta^0\\ 
 D''^{\lambda} \sim \zeta^{\lambda}(\zeta^{\bar{0}}\zeta^0,\zeta^{\bar{4}}\zeta^4)\\
        X''^{\lambda} \sim \zeta^{\lambda}\zeta^{\bar{0}}\zeta^4
        \end{array}   \right)                
\]
implies the existence of a brand new quark $X''$ (of charge -4/3) in a third generation. Though 
$X''$ may be more massive than even the top quark, the consequence at lower energy scales
is that we do not expect the CKM matrix to be quite unitary\footnote{This meshes in with the
observation that an isotriplet couples more strongly with the charged $W$-boson than an
isodoublet and therefore the known decay width of the top quark requires a correspondingly
smaller $V_{tb}$ coupling to $W$.}. Probably the best way to find $X$ is via a high
energy electron-positron collider? Other predictions of the scheme are that heavy leptons 
should be seen as well as unaccompanied (massive?) D-type quarks. If none of these 
signals eventuates then it is back to the drawing board and a reexamination to see if any
of these ideas about property is salvageable or if the disease is terminal.

\section{Force fields}
The most interesting feature of our scheme is the way that gauge fields enter and tie in
with the quantum number assignments.  We note that a flat metric in $X$ is only invariant 
under global SU($N$) unitary rotations of the $N$ attributes. But as soon as we make 
them local or $x$ dependent, so that
\begin{equation}
\zeta^\mu\rightarrow\zeta'^\mu=[\exp(i\Theta(x))]^{\mu\bar{\nu}}\zeta^\nu 
\end{equation}
we find that there is an inconsistency in the transformation rules for the metric; we are
forced to `curve' the space and introduce gauge fields to repair the fault. The
way to do this is to write the generalized event (separation)$^2$ as
\begin{equation}
ds^2 = dX^M dX^N G_{NM}; \quad G_{NM} = {{\cal E}_N}^B{{\cal E}_M}^A \eta_{AB}(-1)^{[B][M]}
\end{equation}
where the metric arises through frame vectors $\cal E$ and the grading is defined in the 
usual way:
$[m]=0, [\mu]=1$. Thus the transformations rules for $G$,
\begin{equation}
G'_{SR}(X') = \left(\frac{\partial X^M}{\partial X'^R}\right)\left(\frac{\partial X^N}{\partial X'^S}\right)
      G_{NM}(X) (-1)^{[S]([R]+[M])},
\end{equation}
under the local rotations of $\zeta$ demand that we introduce components $G_{n\mu},
G_{n\bar{\mu}}$ which have a vectorial character; they should be overall fermionic and 
must somehow involve the gauge field $V$ as this is the communicator of property
across spacetime.  A few moment's reflection (neglecting coupling constants for the present) leads one to the identification 
${{\cal E}_m}^\alpha = -i{V_m}^{\alpha\bar{\nu}}\zeta^\nu$, which is very similar to the
way that the em field makes an appearance in the original Klein-Kaluza model; there is really
very little room for manoeuvre and the appearance is indeed entirely natural: gauge fields transmit property from one place and time to the next so they ought to arise in the spacetime-attribute sector. The only liberty permitted to us is to multiply by polynomials in property
scalars $Z\equiv \zeta^{\bar{\mu}}\zeta^\mu$, since these are gauge invariant and carry no quantum numbers. We might say that inclusion of these polynomials
corresponds to `curving' property space. 

Using that freedom, the only metric which is fully consistent with local SU($N$) gauge transformations is
\begin{equation}
\left(\! \begin{array}{ccc}
        G_{mn} & G_{m\nu} & G_{m\bar{\nu}}\\
        G_{\mu n} & G_{\mu\nu} & G_{\mu \bar{\nu}}\\
        G_{\bar{\mu} n} & G_{\bar{\mu} \nu} & G_{\bar{\mu}\bar{\nu}} 
        \end{array}  \!\right) \!\!=\!\!\
\left(\! \begin{array}{ccc}
        g_{mn}C \!\!+\!\ell^2\bar{\zeta}\{V_m,V_n\}\zeta C'/2 & 
              ~-i\ell^2(\bar{\zeta}V_m)^{\bar{\nu}}C'/2 & ~i\ell^2(V_m\zeta)^\nu C'/2\\
         -i\ell^2(\bar{\zeta}V_n)^{\bar{\mu}}C'/2 & 0 & \ell^2 {\delta_\mu}^\nu C'/2\\
         i\ell^2(V_n\zeta)^\mu C'/2 &  -\ell^2 {\delta_\nu}^\mu C'/2 & 0  
           \end{array}\!  \right).
\end{equation}
Here $C(Z) = 1 + \sum_{n=1}^N c_n Z^n, C'(Z) = 1 + \sum_{n=1}^N c'_n Z^n$ are independent
polynomials of order $N$ in $Z$ which are allowed without destroying
the gauge symmetry. One then readily checks that the rule (6) just corresponds to the usual
gauge transformation: $iV'_m(x') = \exp[i\Theta(x)](iV_m(x) + \partial_m)\exp[-i\Theta(x)]$.
If we just demand subgroup gauge symmetry, we can relax the conditions on the $Z$ 
polynomials and have them invariant under local subgroup rotations, so more property 
curvature coefficients $c_n$ can be entertained. We will come back to this when considering
QCD plus QED and electroweak theory in such a framework.

The procedure from hereon is pretty straightforward\cite{RDPS1}, paying very particular attention
to orders of terms and signs that are due to grading. One first constructs the Super-Ricci
scalar $\cal R$ from the Christoffel symbols
\begin{equation}
2 \Gamma_{MN}{}^K \!\!=\!\!\big[ (-1)^{[M][N]} G_{ML,N}
  +  G_{NL,M}- (-1)^{[L]([M]\!+\![N])} G_{MN,L} \big](-1)^{[L]} G^{LK}
\end{equation}
via the Palatini form
\begin{equation}
{\cal R}=G^{MK}{\cal R}_{KM}= 
   (-1)^{[L]}G^{MK}[(-1)^{[L][M]}{\Gamma_{KL}}^N{\Gamma_{NM}}^L
                - {\Gamma_{KM}}^N{\Gamma_{NL}}^L].
\end{equation} 
Secondly one integrates $\cal R$ over property. This leads to the gravitational and gauge 
field Lagrangian plus a cosmological term\cite{RDPS2}. (As a bonus,  the stress tensor $T_{mn}$ of the gauge fields is automatically incorporated in ${\cal R}_{mn}$ when we 
extract the resulting `equations of motion'.) The coefficients in front of these terms depends 
on the number of properties\cite{RDPS5} and on the property curvature coefficients but they all have the generic form
\begin{equation}
\left( \frac{\ell^2}{2}\right)^{\!\!2(N-1)}\!\!\int\!\! d^N\!\zeta d^N\!\bar{\zeta}\,\sqrt{G..}\,{\cal R}= 
\frac{{\cal A} R^{[g]}}{\ell^2} +{\cal B}{\rm Tr}(F.F) + \frac{\cal C}{\ell^4},
\end{equation}
where $F_{mn}\equiv V_{n,m} - V_{m,n} + i[V_m,V_n]$ and the $N$-dependent coefficients 
${\cal A}, {\cal B}, {\cal C}$ are listed in ref. 7

The matter fields and their Lagrangians are then introduced,
\begin{eqnarray}
{\cal L}_\phi&=&\int\!\! d^N\!\zeta d^N\!\bar{\zeta}\,\sqrt{-G..}\,\,
  G^{MN}\partial_N\Phi\partial_M\Phi; \\
{\cal L}_\psi&=&\int\!\! d^N\!\zeta d^N\!\bar{\zeta}\,\sqrt{-G..}\,\,
 \bar{\Psi}\,i\Gamma^A{E_A}^M\partial_M\Psi.
\end{eqnarray}
The gauge and gravitational interactions of the component fields ($\phi,\psi$) 
then just fall out, but these sometimes require wave function
renormalizations due to influence of the property curvature coefficients 
$c_n$ --- coefficients which are absent in flat space. The key point is that the gauge fields
couple correctly to the matter fields through the vielbein term
\[
{E_A}^M\partial _M \supset {e_a}^m [\partial_m+i{(V_m\zeta)}^\mu\,\partial_\mu
                  -i(\bar{\zeta}V_m)^{\bar{\mu}}\,\partial_{\bar{\mu}}],
\]
so the property derivative is compensated by a further property coordinate attached to
the gauge field $V$; this is our version of covariant differentiation. Incidentally
I ought to declare that  such complicated calculations were originally
carried using an algebraic computer package devised by Paul Stack and, after time
consuming computation, they always produced gauge- and coordinate-invariant results. Knowing this always happened, we have since been able to find a shorter analytic way of 
picking out the correct terms in (10)-(12) by a procedure which can be generalized to any 
number of attributes and dispense with Mathematica.
Finally, to (11) and (12) we may add the renormalizable super-Yukawa self 
interactions $\Psi\Phi\Psi$ and $V(\Phi) \simeq \Phi^4$ in the usual manner, 
with the aim of generating a mass term through the expectation values held in 
the chargeless fields within $\langle\Phi\rangle$.

Before moving on, three comments about the fermion fields deserve particular mention.
Firstly the adjoint field $\overline{\Psi}$ has to be carefully defined with appropriate signs\cite{RDPS4} in property space to produce a series of terms $\bar{\psi}\psi$, after 
integrating over $\zeta$.
Secondly, $\bar{\zeta}\psi$ and their charge conjugates $\psi^{(c)}\zeta$ both
appear in the full expansion of $\Psi(\zeta,\bar{\zeta})$ and they simply lead to a doubling
of the eventual answers; thus we can simplify calculations by `halving' the expansion of 
$\Psi$ to ${\Psi} \supset \bar{\zeta}\psi$ terms. Thirdly and  intriguingly, we have to extend
the concept of Dirac $\gamma$ matrices  to super $\Gamma$ matrices, such that
$(\Gamma^A P_A)^2 = \eta^{AB}P_BP_A$. In spacetime we get the standard 
$\Gamma^a =\gamma^a$ with $\{\gamma^a,\gamma^b\}=2\eta^{ab}$, but in the 
property sector one needs to ensure that the `square-rooted' $\Gamma^\alpha$ 
are {\em fermionic} and obey
\begin{equation}
[\Gamma^\alpha,\Gamma^\beta]=[\Gamma^{\bar{\alpha}},\Gamma^{\bar{\beta}}]=0,
\qquad [\Gamma^\alpha, \Gamma^{\bar{\beta}}]=2\eta^{\alpha\bar{\beta}}=
2{\delta_\beta}^\alpha.
\end{equation}
In the same way that Dirac introduced 4$\times 4$ matrices and made novel use of the Clifford algebra for spacetime, we must do something similar for property space. We can arrange for
the commutators (13) to be satisfied by augmenting property space with auxiliary coordinates
$\theta^\alpha$, setting $\Gamma^\alpha\equiv\sigma_+\theta^\alpha$, 
$\Gamma^{\bar{\alpha}}\equiv\sigma_-\partial/\partial\theta^\alpha$, and making sure that 
$\Psi$ is multiplied by the projected singlet 
$\Theta\equiv (1+\sigma_3)\theta^1\theta^2\cdots \theta^N/2$, 
over which one eventually integrates\footnote{Of course the adjoint $\overline{\Psi}$ 
contains the conjugate singlet $\bar{\Theta} =(1+\sigma_3)\theta^{\bar{N}}\cdots
\theta^{\bar{2}}\theta^{\bar{1}}/2$.}. There are probably less extravagant ways of 
doing this.

\section{Electric and Chromic Relativity}
To see how all this works out consider QED and QCD which involve one attribute
called electricity plus three `chromicity' properties (commonly termed red, green, blue).
Thus we confine ourselves to coordinates $\zeta^1$ to $\zeta^4$ and combine both
chiralities in Dirac fields since those interactions are blind to parity. Because we are
confining ourselves to U(1)$\times$SU(3) we are dealing with two sets of gauge
fields within the fuller SU(4): the em field $A$ and the gluon fields $B$, having coupling
constants $e$ and $f$ respectively. One identifies the frame vectors
${{\cal E}_m}^\kappa = -i({fB_m-eA_m/3)}^{\kappa\bar{\iota}}\zeta^{\iota},\,\,
{{\cal E}_m}^4=ieA_m$, leading to the basic metric elements
\begin{equation}
G_{m4}= i\ell^2\zeta^{\bar{4}}eA_mC'/2,\quad G_{m\iota}=
i\ell^2[\zeta^{\bar{\iota}}eA_m/3 -\zeta^{\bar{\kappa}}f{B_m}^{\kappa\bar{\iota}}]C'/2,
\end{equation}
which may be multiplied by polynomials in two {\em distinct} invariants 
$\zeta^{\bar{\kappa}\zeta^\kappa}$ and $\zeta^{\bar{4}}\zeta^4$. I should point
out that it is the interactions (14) which actually determine the charge and colour 
assignments stated in (2) and (3). Also the coupling {\em must} accompany the
gauge fields in order to produce the correct interactions with matter fields.

To simplify the subsequent argument about the resulting interactions I will assume that 
the property curvature polynomials are common to spacetime \& property space:
\begin{equation}
C=C'= 1 +\cdots +c_e(\zeta^{\bar{4}}\zeta^4)(\zeta^{\bar{\kappa}}\zeta^\kappa)^2
+ c_f(\zeta^{\bar{\kappa}}\zeta^\kappa)^3.
\end{equation}
As we are dealing with four properties, we find that the Berezinian\cite{B,deW}  is
$\sqrt{-G..} = (2/\ell^2)^4\sqrt{-g..} C^{-2}$. A careful analytical calculation 
shows that the super-Ricci scalar contains the following gauge field combination:
\begin{eqnarray}
{\cal R}\sqrt{-G..}\supset&& [1-3c_e(\zeta^{\bar{4}}\zeta^4)(\zeta^{\bar{\kappa}}\zeta^\kappa)^2
 -3c_f(\zeta^{\bar{\kappa}}\zeta^\kappa)^3+\cdots ]\nonumber \\
 &&\sqrt{-g..}g^{km}g^{ln}[4e^2\zeta^{\bar{4}}F_{kl}F_{mn}\zeta^4/3 +
  f^2\zeta^{\bar{\kappa}}(E_{kl}E_{mn})^{\kappa\bar{\iota}}\zeta^\iota],
\end{eqnarray}
where $F_{mn}=A_{n,m}-A_{m,n}$ and $E_{mn}=B_{n,m}-B_{m,n}+if[B_n,B_m]$
are the standard `curls' of the electromagnetic and gluon fields. The last step
is to integrate over the four properties. Including appropriate scaling factors of 
$\ell^2$ one gets
\begin{equation}
\int(d^4\zeta d^4\bar{\zeta})\,{\cal R}\sqrt{-G..}\supset (-12\sqrt{-g..}/\ell^2)[4c_fe^2F.F 
+ c_ef^2{\rm Tr}(E.E)].
\end{equation}
Last but not least we must ensure gravitational universality; so we have to set 
$c_ef^2 = 4c_fe^2$, which is perfectly feasible without demanding equality 
of the colour and electromagnetic couplings. If we relax the assumption that 
$C=C'$, it is even easier to ensure universality of Newton's constant $G_N$. 

The colour and electromagnetic interactions of the matter fields $\Psi,\Phi$ emerge from
(11) and (12) exactly as expected. See ref. 8. I shall not delve into that because the 
story is not quite complete and is therefore likely to be misleading:  we have neglected
neutrinicity (the fifth property $\zeta^0$) so the ensuing generations are not the 
physical ones, as sketched in section 3. To correct for this we must turn to the leptons.

\section{Electroweak Relativity}
The application of our scheme to the original electroweak model\cite{G,W,S} of leptons requires
an interesting extension of previous work\cite{RDPS3}  and leads to an intriguing 
prediction about the weak mixing angle, not to mention the prediction of two leptonic 
generations. The fact that the weak isospin and hypercharge assignments of the leptons 
change with chirality obliges us to invoke distinct properties $\zeta_L$ and $\zeta_R$ for 
left- and right-handed leptons respectively to which the gauge fields latch on (through the 
frame vectors). The full SU(4) gauge field $V^{\mu\bar{\nu}}$, acting on the pair of doublets
$(\zeta_L^0,\zeta_L^4,\zeta_R^0,\zeta_R^4)$ is not needed; only the restricted
SU(2)$_L\times $U(1) rotations demand attention. Thus we reinterpret 
$V_m = L_m + R_m$, 
with
\begin{equation}
L_m = (g{\mathbf W}_m.{\mathbf \tau} - g'B_m)/2 \qquad R_m= g'B_m(\tau_3-1)/2,
\end{equation}
possessing the standard weak hypercharge assignments
\begin{equation}
Y(\zeta_L^0,\zeta_L^4,\zeta_R^0,\zeta_R^4) = (-1,-1,0,-2).
\end{equation}
It must also be understood that $L$ is to be associated with the left property derivative
$\partial/\partial \zeta_L$ and $R$ is to be associated with the right property 
derivative $\partial/\partial \zeta_R$; $g$ and $g'$ are the usual coupling constants
tied to the weak triplet ${\mathbf W}$ and weak singlet hypercharge $B$ respectively.

It is sufficiently general for our purposes to take the polynomial property curvatures $C$
and $C'$ to  be equal and direct products of quadratic left- and right-handed polynomials:
\begin{equation}
C = C_R C_L = [1+c_RZ_R + c_{RR}Z_R^2][1+c_L+c_{LL}Z_L^2]; 
  \,Z_R\equiv \bar{\zeta}_R\zeta_R,\, Z_L\equiv \bar{\zeta}_L\zeta_L.
\end{equation}
These enter in the metric components:
\begin{eqnarray}
G_{m\zeta_L}&=&-i\ell^2\bar{\zeta}_L L_mC/2;\qquad 
G_{m\zeta_R}=-i\ell^2\bar{\zeta}_R R_mC/2,\\
G_{\zeta_L \bar{\zeta}_L}&=&G_{\zeta_R \bar{\zeta}_R}=\ell^2 C/2,
\quad G_{\zeta_L\zeta_R}=G_{\bar{\zeta}_L\bar{\zeta_R}}
=G_{\zeta_L \bar{\zeta}_R}=G_{\zeta_R \bar{\zeta}_L}=0.
\end{eqnarray}
The remaining metric element reads
\begin{equation}
G_{mn}=C[g_{mn} + (\rm gauge~field~terms)].
\end{equation}
Factorisability of $C$ simplifies the calculations enormously when we integrate
over the whole eight properties: 
$\int d^2\zeta_Rd^2\bar{\zeta}_R\,d^2\zeta_Ld^2\bar{\zeta}_L$.

The various contributions to the super-Ricci scalar drop out as follows, bearing
in mind that  $\sqrt{-G..} = (2/\ell^2)^4\sqrt{-g..}(C_RC_L)^{-3}$. There are three terms:
\begin{equation}
\int d^2\zeta_R..d^2\bar{\zeta}_L\ \sqrt{-G..}{\cal R}\supset 36\sqrt{-g..}(2/\ell^2)^4\,
     R^{[g]} (2c_R^2-c_{RR})(2c_L^2-c_{LL}),
\end{equation}
\begin{eqnarray}
\int\!\!d^2\zeta_R..d^2\bar{\zeta}_L\,\sqrt{G..}{\cal R}&\!\supset&\! -\frac{3}{2}\sqrt{g..}
\left(\!\frac{2}{\ell^2}\!\right)^{\!\!3}\,
     \!\!\!\left[c_L(3c_R^2\!-\!2c_{RR})(g^2{\mathbf W}_{mn}.{\mathbf W}^{mn}\!\!\!+\!
     g'^2B_{mn}B^{mn})\right.\nonumber\\
     & & \qquad\qquad\qquad \left. +g'^22c_R(3c_L^2-2c_{LL})B_{mn}B^{mn}\right],
\end{eqnarray}
\begin{equation}
\int d^2\zeta_R..d^2\bar{\zeta}_L\,\sqrt{G..}{\cal R}\supset 12\sqrt{g..}(2/\ell^2)^5\,
     [(4c_Lc_{LL}\!\!-\!5c_L^2)(2c_R^2\!-\!c_{RR})+(L\leftrightarrow R)],
\end{equation}
where $W_{mn}\equiv W_{n,m}-W_{m,n}+ig[W_n,W_m]$ and 
$B_{mn}\equiv B_{n,m}-B_{m,n}$.  The full answer is the sum of (24)-(26).

Universality of gravity at the semiclassical level anyway (and the correct normalization 
of the gauge fields) is  guaranteed when we set
\[
c_L(3c_R^2-2c_{RR})(g^2-g'^2) = 2c_R(3c_L^2-2c_{LL})g'^2,
\]
which is readily arranged. But much more intriguing is the fact that if the property 
curvature is parity conserving so that $c_R=c_L=c, c_{RR}=c_{LL}=c_2$ and implying that 
all parity violation comes from the gauge fields in the frame vectors, then $g^2=3g'^2$.
Thus the weak angle reduces to 30$^o$. It makes good sense because the
property curvature $C$ polynomial accompanies
the gravitational field and, as far as we know, gravity does not know the left hand
from the right. So this restriction seems very natural and the value of the weak
angle is a consequence of gravitational universality in this framework; it is not a result of 
invoking a higher group or anomaly cancellation, as some other analyses\cite{DK,I} 
would have.

Turning to the matter fields, we can reduce the number of components by invoking 
selfduality\footnote{SU(2) duality, indicated by $\times$, stipulates that
$1^\times = Z^2/2, (\zeta^\mu)^\times\! =\! (\zeta^\mu)Z, Z^\times=Z, 
(\eta_{\mu\nu}\zeta^\mu\zeta^\nu)^\times = -\eta_{\mu\nu}\zeta^\mu\zeta^\nu$, where
$Z=\zeta^{\bar{\mu}}\zeta^\mu$. Vice versa, and likewise for the hermitian 
conjugate combinations. Thus the selfdual combinations are $(1 + Z^2/2), Z$ and 
$\zeta(1+Z)$ with $\eta_{\mu\nu}\zeta^\mu\zeta^\nu\rightarrow 0$. We apply this 
separately to left and right leptonic properties in the following equations,
corresponding to the subgroup $SU(2)_L\times SU(2)_R$.}
(corresponding to symmetry about the cross diagonal in the superfield expansions).
Ignoring the charge conjugate terms, which simply double the results below, two 
fermion generations, $\psi$ and $\psi'$ arise from expanding $\Psi$. Using the shorthand
symbols $Z_L\equiv\bar{\zeta_L}\zeta_L, Z_R\equiv\bar{\zeta_R}\zeta_R$  as in (20), we get
\begin{eqnarray}
2\Psi&=&\bar{\zeta}_L[\psi_L(1+Z_R^2/2)+\psi'_L  Z_R](1+Z_L)+(L\leftrightarrow R),\\
2\overline{\Psi}&=&[\overline{\psi_L}(1+Z_R^2/2)+
      \overline{\psi'_L}Z_R]\zeta_L(1+Z_L)+(L\leftrightarrow R).
\end{eqnarray}
Since chirality ensures that  $\overline{\psi_L}\psi_L=\overline{\psi_R}\psi_R=0$, we
find that a mass term arising from the product $\overline{\Psi}\Psi$ has insufficient powers
of $\zeta$ to give a nonzero answer; thus a {\em mass term vanishes identically} and this is 
a good  thing because it indicates that we need to couple fermions to bosons before one can
generate mass. The kinetic term is fine however; in flat space,
\begin{equation}
- \int\!\!d^2\zeta_R..d^2\bar{\zeta}_L\overline{\Psi}i\gamma.\partial\Psi =
\overline{\psi_L} i\gamma.\partial\psi_L + \overline{\psi'_L} i\gamma.\partial\psi'_L 
+ (L\leftrightarrow R).
\end{equation}

Regarding the bosons, we recall that the selfdual combinations are $(1+Z^2/2)$ 
and $Z$ with $\zeta\zeta\rightarrow 0$, separately for left- and right-handed properties.
Hence the fully selfdual, hermitian superBose field $\Phi$ is 
\begin{eqnarray}
2\Phi &=& \varphi(1+Z_L^2/2)(1+Z_R^2/2)+\varphi'Z_LZ_R + 
    \Lambda Z_L(1+Z_R^2/2)+P Z_R(1+Z_L^2/2)+ \nonumber \\
 & & \quad[\bar{\zeta}_R\phi\zeta_L +\bar{\zeta}_L\phi^\dag\zeta_R +
       \phi'\zeta_R\zeta_L + \bar{\zeta}_L\bar{\zeta_R}\phi'^\dag] (1+Z_L)(1+Z_R). 
\end{eqnarray}
If we further restrict ourselves to fields of even parity under the operation 
$\zeta_R\leftrightarrow \zeta_L$, we find $\Lambda=P\equiv\chi/\sqrt{2}, 
\varphi'=0, \phi=\phi'$,
so the expansion (30) reduces to
 \begin{eqnarray}
2\Phi &=& \varphi(!+Z_L^2/2)(1+Z_R^2/2)+\varphi'Z_LZ_R + 
    \chi [Z_L(1+Z_R^2/2)+ Z_R(1+Z_L^2/2)]/\sqrt{2} \nonumber \\
 & & \qquad+[\bar{\zeta}_R\phi\zeta_L +\bar{\zeta}_L\phi^\dag\zeta_R](1+Z_L)(1+Z_R). 
\end{eqnarray}
The normalization factors have been concocted so that
\begin{equation}
- \int\!\!d^2\zeta_R..d^2\bar{\zeta}_L \,\Phi^2= -\varphi^2-\varphi'^2-\chi^2 + {\rm Tr}(\phi^2).
\end{equation}
In (31) the quartet $\phi^{\mu\bar{\nu}}=
     (\phi_oI+{\mathbf \phi}.{\mathbf \tau})^{\mu\bar{\nu}}/\sqrt{2}$ consists of a singlet
and a triplet. The quantum numbers $I_{3L},Y,Q=I_{3L}+Y/2$ of the components read:
\begin{eqnarray*}
Y(\varphi,\varphi',\chi)&=&(0,0,0);\quad I_{3L}(\varphi,\varphi',\chi)=(0,0,0);\quad
Q(\varphi,\varphi',\chi)=(0,0,0);\\
Y(\phi^{0\bar{0}},\phi^{0\bar{4}},\phi^{4\bar{0}},\phi^{4\bar{4}})&=&(1,1,-1,-1);\qquad
2I_{3L}(\phi^{0\bar{0}},\phi^{0\bar{4}},\phi^{4\bar{0}},\phi^{4\bar{4}})=(-1,1,-1,1);\qquad\\
Q(\phi^{0\bar{0}},\phi^{0\bar{4}},\phi^{4\bar{0}},\phi^{4\bar{4}})&=&(0,1,-1,0).
\end{eqnarray*}
The Higgs boson will be associated with $\phi_0+\phi_3$, as we will presently discover.
For that identification we need to consider the super-Yukawa and gauge field interactions
in flat spacetime, before we curve spacetime with gravity.

With $L$ and $R$ gauge fields defined in (18), the vielbeins which correspond to the metric elements (21) - (23) are:
\begin{equation}
\left( \begin{array}{ccc}
       {E_a}^m & {E_a}^\mu & {E_a}^{\bar{\mu}}\\
       {E_\alpha}^m & {E_\alpha}^\mu & {E_\alpha}^{\bar{\mu}}\\
       {E_{\bar{\alpha}}}^m & {E_{\bar{\alpha}}}^\mu & {E_{\bar{\alpha}}}^{\bar{\mu}}
        \end{array} \right) =  \frac{1}{\sqrt{C}}
        \left( \begin{array}{ccc}
        {e_a}^m & i[(L_a \zeta_L) + (R_a \zeta_R)]^\mu & 
                      - i[(\bar{\zeta}_L L_a)+ (\bar{\zeta}_R R_a)]^{\bar{\mu}}\\
        0 & {\delta_\alpha}^\mu & 0\\
        0 & 0 & {\delta_{\bar{\alpha}}}^{\bar{\mu}}
        \end{array}   \right).  
\end{equation}
Thus the fermion kinetic energy can be written in the form 
$\overline{\Psi}i\Gamma^A D_A \Psi$, where
\[
D_A =  {E_A}^M\partial_M = {E_A}^m\partial_m + {E_A}^\mu\partial_\mu + 
            {E_A}^{\bar{\mu}}\partial_{\bar{\mu}}
\]
acts like a covariant derivative. Let $V$ serve as a generic gauge field; the action of 
$i\gamma^a D_a$ on 
$f(Z)(\bar{\zeta}\psi)$ is to give $f(Z)\bar{\zeta}\gamma.(i\partial_a+V_a)\psi$
and on $f(Z)(\bar{\psi}\zeta)$ is to give $f(Z)(i\partial_a + V_a)\bar{\psi}\gamma^a\zeta$.
So when we integrate over property we end up precisely with the usual
gauge field interaction $\bar{\psi}\gamma^a(i\partial_a+V_a)\psi$ for each of
the two generations, which in the leptonic case translates into
\[
\overline{\psi_L}\gamma^a(i\partial_a+L_a)\psi_L 
+\overline{\psi_R}\gamma^a(i\partial_a+R_a)\psi_R + (\psi\rightarrow \psi').
\]
This is unsurprising; interpreting $(\psi^0,\psi^4)= (\nu,l)$, one ends up with the 
standard
\begin{eqnarray}
{\cal L}_\psi &=&\bar{l}\gamma.(i\partial -eA)l+\bar{\nu}i\gamma.\partial\nu 
+ \frac{e}{\sqrt{2}\sin\theta}[\overline{\nu_L}\gamma.W^+l_L+\overline{l_L}\gamma.W^-\nu_L]
\nonumber \\
& & +\frac{e}{\sin 2\theta}(\overline{\nu_L}\gamma.Z\nu_L)+
e\tan\theta(\overline{l_R}\gamma.Z l_R)-e\cot 2\theta (\overline{l_L}\gamma.Z l_L)\nonumber\\
&&+\,\,(l,\nu)\rightarrow (l',\nu'),
\end{eqnarray}
where $\cos\theta=g/\sqrt{g^2+g'^2},\,\,\sin\theta=g'/\sqrt{g^2+g'^2},\,\, e=gg'/\sqrt{g^2+g'^2}$.
(34) simplifies to a considerable extent when $\theta=30^o$,  as indicated by gravitational universality, because the $Z$ field then interacts purely axially with the charged lepton, in
contrast to the purely vectorial electromagnetic field.

But when we come to the bosons we discover something new. Acting with the covariant
derivative on the Bose superfield, 
\[
D_a\Phi=[{E_a}^m\partial_m+{E_a}^\mu\partial_\mu+{E_a}^{\bar{\mu}}\partial_{\bar{\mu}}]\Phi
=[\partial_a + i(V_a\zeta)^\mu\partial_\mu-i(\bar{\zeta}V_a)^{\bar{\mu}}\partial_{\bar{\mu}}]\Phi.
\]
Referring to eq. (31) we obtain
\begin{equation}
2D\Phi.D\Phi=(1+2Z_L)(1+2Z_R)[\bar{\zeta}_R\{\partial\phi+i(\phi L-R\phi)\}\zeta_L
         \bar{\zeta}_L\{\partial\phi+i(\phi R-L\phi)\zeta_R\}]
\end{equation}
plus terms which disappear when integrated over property. If we concentrate on the uncharged 
fields held in the quartet $\phi$, viz. $\phi^{0\bar{0}}~\&~\phi^{4\bar{4}}$, that occur on the
diagonal (or equivalently $\phi_0~\&~ \phi_3)$, we find that 
\begin{eqnarray*}
2{\rm Tr}[(\phi R-L\phi)(\phi L-R\phi)]&\rightarrow&\frac{1}{2}g^2W^+W^-(\phi_+^2+\phi_-^2)
      +\frac{1}{4}\phi_+^2(gW_3-g'B)^2\\
      && + \frac{1}{4}\phi_-^2(gW_3+g'B)^2 -g'^2\phi_-^2; \,\,\phi_\pm\equiv \phi_0\pm\phi_3.
\end{eqnarray*}
In order to recover the standard vector meson masses, we must therefore take
\[
\langle\phi_-\rangle = 0 {\rm ~or~} \langle\phi_0\rangle = \langle\phi_3\rangle, {\rm ~and~}
\langle \phi_+\rangle = v,
\]
for the expectation values, whereupon
\begin{eqnarray}
\langle 2{\rm Tr}[(\phi R-L\phi)(\phi L-R\phi)]\rangle &\rightarrow &
      \frac{1}{2}v^2g^2W^+W^- +\frac{1}{4} v^2(g^2+g'^2)Z^2\nonumber \\
      &=&\frac{e^2v^2}{2\sin^2\theta}W^+W^- +\frac{e^2v^2}{\sin^2 2\theta} Z^2.
\end{eqnarray}
All is as it should be and the em field $A$ remains massless.

Given these expectation values, we turn to the Yukawa interaction of the super-Bose 
field $\Phi$ with the  super-Fermion field $\Psi$. Before launching into this we need to remind ourselves that in order to get masses for leptons {\em as well as neutrinos}, we have to
consider the Higgs doublet $H$ as well as its doublet counterpart $i\tau_2H^*$. In our 
context it means that we have to consider $\phi$ as well $\tau_2\phi^*\tau_2$. Since we will 
be integrating over the $\zeta$ and the fermion pieces involve $\bar{\zeta}_R\zeta_L$ or
${\bar{\zeta}}_L\zeta_R$, we need to pick out matching bose pieces. 
Using the acceptable combination\footnote{With such
a combination, ${\rm Tr}\,\,\hat{\phi}^2 =( c_l^2+s_l^2)(\phi_0^2+ \mathbf{\phi}^2) +
2c_ls_l(\phi_0^2-\mathbf{\phi}^2)$. So taking expectation values,
${\rm Tr}\,\,\langle\hat{\phi}\rangle^2= 2(c_l^2+s_l^2)v^2\rightarrow 2v^2$ if we
interpret $c_l\equiv \cos\theta_l,\,s_l\equiv \sin\theta_l$.}
$\hat{\phi}= c_l\tau_2\phi^*\tau_2 + s_l\phi$,  in place of $\phi$,
we then find that
\begin{eqnarray*}
-8\sqrt{2}\overline{\Psi}\Phi\Psi  &\supset &
(\bar{\zeta}_R\hat{\phi}\zeta_L+\bar{\zeta}_L\hat{\phi}^\dag\zeta_R)(1+2Z_L)(1+2Z_R).\\
& & [\bar{\zeta}_L\zeta_R(\overline{\psi_R}+Z_R\overline{\psi'_R})(\psi_L+Z_L\psi'_L)
        + (R \leftrightarrow L)].
\end{eqnarray*}
Consequently, integrating over property produces a mixture of the two generations:
\[
-16\int d^2\zeta_R..d^2\bar{\zeta}_L \,\overline{\Psi}\Phi\Psi =
 (2\bar{\psi}+\bar{\psi}')\hat{\phi}(2\psi+\psi')\equiv 
  5\overline{\hat{\psi}}\hat{\phi}\hat{\psi}.
\]
Taking expectation values of $\Phi$ to generate a fermionic mass term, and
recalling that $\langle\phi_-\rangle=0$, the Yukawa term (including a coupling
constant $\mathfrak{g}$) reduces to
\begin{equation}
5\mathfrak{g}(\overline{\nu_l}\,,\,\bar{l})\left(\begin{array}{cc} 
                                        c_l\langle\phi_+\rangle & 0\\
                                        0 & s_l\langle\phi_+\rangle \end{array}\right)
                                       \left(\begin{array}{c}  \nu_l\\ l\end{array}\right)
                     =5v\mathfrak{g}(c_l \overline{\nu_l}\nu_l + s_l \bar{l}l).                  
\end{equation}
The other mixture $\check{\psi}=(-\psi+2\psi')/\sqrt{5}$ does not acquire a mass in this 
model. If we were to stretch credulity and pretend we have a decent model for leptons we 
would be inclined to associate $\hat{\psi}$ with the muonic doublet and $\check{\psi}$
with the electronic one; but all this is academic: we really need the three colour properties
to corall the known leptonic generations.

The last thing to consider is the effect of spacetime curvature (through ${e_m}^a$ or
$g_{mn}$) and of property curvature $C(Z)$ on the above results. The effect of $e$ is very simple: it just serves to make the interactions generally covariant and we have nothing 
more to add to that. The effect of $C$ enters through the Berezinian 
\[
\sqrt{G..}\!=\!\sqrt{g..}(2/\ell^2)^4C^{-2}\!\propto\! 
(1-2c_R-2c_{RR}Z_{RR}+3c_R^2Z_R^2)(1-2c_L-2c_{LL}Z_{LL}+3c_L^2Z_L^2).
\]
It is subtler and causes mixing as well as wavefunction renormalization. 
To see what happens, consider the kinetic term of the fermions and simplify the
argument by assuming the property curvature is blind to parity as we did before
to recover a weak mixing angle of 30$^o$. In that case, using the expanded
\begin{eqnarray}
\sqrt{G..}\!&=&\!\sqrt{g..}(2/\ell^2)^4[1-2c(Z_R+Z_L)+4c^2Z_RZ_L+\\
& &\, (3c^2-2c_2)\{Z_R^2(1-2cZ_L)\!+\!Z_L^2(1-2cZ_R)\}+\{(3c^2-2c_2)Z_LZ_R\}^2],
\end{eqnarray}.
we obtain, after $\zeta$ integration, the kinetic term ($D\equiv\partial-iV$),
\[
\sqrt{g..}[(1-c)\{\overline{\psi}i\gamma.D\psi + \overline{\psi'}i\gamma.D\psi'
 -2c(\overline{\psi}i\gamma.D\psi'+\overline{\psi'}i\gamma.D\psi)\}
+ c(c^2+2c_2)\bar{\psi}i\gamma.D\psi]
\]
which reduces to (29) when $c\rightarrow 0$. Thus the curving of property engenders
source field mixing and wavefunction renormalization, without affecting the coupling of
the gauge field $V$ configuration. Similar conclusions apply to the Bose sector.

\section{Generalizations and Conclusions}
I have outlined the main consequences of a mathematical scheme for handling
the `when-where-what' of events by an enlarged coordinate backdrop, part being
commuting (spacetime) and part anticommuting (property).  It automatically
produces a finite number of generations of elementary particles and 
provides a framework that unifies
gravity with the other forces of nature. We treated the case of strong and
electromagnetic interactions SU(3)$\times$U(1) corresponding to
three chromicity and one electricity property, making for a total of four P-conserving
properties. Then we considered augmenting these by neutrinicity to describe electroweak
theory and there we found the need to distinguish between left and right leptons.
Thus the minimal number of properties $\zeta$ for encompassing the known forces is 5 (or 7 
if we double up for leptonic handedness). The full story requires the use of them all and
I admit to not having properly tackled that yet. It is a daunting business as you 
have seen from the calculations presented earlier. We went on to show that if the property 
curvature coefficients respect parity -- which befits gravity at any rate -- the weak mixing angle 
must equal 30$^o$ to guarantee gravitational universality. Also we proved that
the simplest generalization of the standard electroweak model resulted in {\em two} lepton
generations, one massive and one massless, and in addition
was able to reproduce what we know about vector masses. We remain nonplussed 
as to how to constrain the coefficients $c_n$ which curve property and we are still
searching for a principle that will do the job.

To fully handle the complete SU(3)$\times$SU(2)$_L\times$U(1) gauge group, 
rather than bits and pieces, will require more calculational acrobatics and 
is left for future research. Suffice it to say that we have come across obstacles 
and have so far circumvented them all. Whether we will be able to overcome 
looming problems is quite another matter: it may well turn out that the
predictions which emerge will not be able to withstand experimental scrutiny. 
We have set our sights on reproducing the standard model, with the particle
generations automatically catered for. If this succeeds, one can look farther afield,
seeing as we have barely scratched the surface of the scheme. A left-right symmetric
picture beckons; sterile states that do not interact with the
basic constituents exist aplenty in the expansions of $\Psi$ and $\Phi$
and, if we think fancifully, may have connections to dark matter; finally the
quantization via BRST seems to find a natural place in our framework
since it introduces anticommuting scalar variables attached to the ghost fields, 
leading to an Sp(2) translation group. On a more cautionary note,
the future may judge the entire approach as being completely
misguided; after all, just one ugly fact can slay a beautiful hypothesis. The history 
of physics is littered with such failures. If so, the present scheme can be buried
with lots of other valiant attempts in the graveyard of failed theories, but its ghost may
linger awhile.

\section*{Acknowledgements}
I wish to express my thanks to Dr Paul Stack for his computational wizardry in Mathematica 
and his numerous accurate contributions to this topic. If there are any errors in this paper 
they are entirely my own. Also I am indebted to Dr Peter Jarvis for his insights and encouragement over the years. Finally I would like to record the generous support I
have received from the organizers of this splendid meeting.


\begin{thebibliography}{99}

\bibitem{RDDyson}
R.~Delbourgo, {\em Int. J. Mod. Phys.} {\bf 28A}, 1330051 (2013).

\bibitem{GG}
H.~Georgi and S.~Glashow, {\em Phys. Rev. Lett.}, {\bf 32B}, 438 (1974). 

\bibitem{FM}
H.~Fritzsch and P.~Minkowski,  {\em Ann. Phys.} {\bf 93}, 193  (1975).

\bibitem{DJW}
R.~Delbourgo, P.~D. Jarvis and R.~C.Warner {\em Aus. J. Phys.} {\bf 44}, 135 (1991).

\bibitem{RDPS1}
R.~Delbourgo and P.~D. Stack, {\em Int. J. Mod. Phys.}, {\bf 29A}, 50023 (2014)

\bibitem{RDPS2}
P.~D. Stack and R.~Delbourgo, {\em Int. J. Mod. Phys.}, {\bf 30A}, 1550005 (2015).

\bibitem{RDPS5}
R.~Delbourgo and P.~D. Stack, {\em Mod. Phys. Lett.} {\bf 31A},1650019 (2016).

\bibitem{RDPS4}
P.~D. Stack and R.~Delbourgo, {\em Int. J. Mod. Phys.} {\bf 30A}, 1550211 (2015).

\bibitem{B}
F.~A. Berezin, General Concept of Quantization, {\em Comm. Math. Phys,} {\bf 40}, 153 (1975).

\bibitem{deW}
B.~S. DeWitt, {\em Phys. Rept.} {\bf 19}, 295 (1975).

\bibitem{G}
S.~L. Glashow, {\em Nucl. Phys.} {\bf 22}, 579 (1961).

\bibitem{W}
S.~Weinberg, {\em Phys. Rev. Lett.} {\bf 19}, 1264 (1967).

\bibitem{S}
A.~Salam, {\em Eighth Nobel Symposium}, ed. N.~Svartholm, Almquist and Wiksell (1968).

\bibitem{RDPS3}
R.~Delbourgo and P.~D. Stack, {\em Int. J. Mod. Phys.} {\bf 30A}, 1550095 (2015).

\bibitem{DK}
S.~Dimopoulos and D.~E. Kaplan, {\em Phys. Lett.} {\bf B531}, 127 (2002).

\bibitem{I}
L.~E. Ibanez, {\em Phys. Lett.}  {\bf B303}, 65 (1993).

\end{thebibliography}
\end{document}